\documentclass[a4paper,11pt]{article}
\usepackage[utf8]{inputenc}
\usepackage[T1]{fontenc}
\usepackage[english]{babel}
\usepackage[width=160mm,height=247mm]{geometry}
\usepackage{amsmath,amsfonts,amssymb,bm}
\usepackage{lmodern,graphicx,float,indentfirst,physics,xcolor,cite}%
\usepackage[squaren]{SIunits}
\usepackage[textfont={footnotesize},labelfont={color=blue,bf,sf},labelsep=endash]{caption}
\usepackage[position=top,labelfont={color=blue,bf,sf}]{subfig}
\usepackage[colorlinks,urlcolor=blue,linkcolor=blue,citecolor=blue]{hyperref}

\usepackage[affil-sl]{authblk}
\begin{document}

\setlength{\affilsep}{1em}
\renewcommand\Authfont{\normalsize\bf}
\renewcommand\Affilfont{\it\normalsize}
\title{\textbf{\Large Goos-Hänchen Shifts in Graphene with Spatially\\
Modulated Potential}}
\author[a]{Abdellatif Kamal\thanks{\href{mailto:kamal.a@ucd.ac.ma}{\sf kamal.a@ucd.ac.ma}}}
\author[a]{Ahmed Jellal\thanks{\href{mailto:a.jellal@ucd.ac.ma}{\sf a.jellal@ucd.ac.ma}}}
\affil[a]{Laboratory of Theoretical Physics, Faculty of Sciences, Choua\"ib Doukkali University,}
\affil[ ]{PO Box 20, 24000 El Jadida, Morocco}
\date{}
\providecommand{\pacs}[1]{\noindent PACS numbers: #1\\}
\providecommand{\keywords}[1]{\noindent Keywords: #1}
\hypersetup{pageanchor=false}
\begin{titlepage}
    \maketitle
    \thispagestyle{empty}
    \vspace{3cm}
    \begin{abstract}
        Using the energy spectrum of 
        graphene with spatially modulated potential, we study the Goos-Hänshen shifts near extra Dirac points located at finite energy $\varepsilon=m\pi$, with $m$ integer. 
        On both sides of such points, we show 
         that the Goos-Hänshen shifts can be negative as well as positive
         under various conditions. %
         It is found that 
        the shifts are strongly depending on the effects of the incident energy, potential height, incident angle and  width 
        of central region of unit cell. Such effects tell us that such width can be used to tune 
        the shifts at the extra Dirac points.
    \end{abstract}

    \vspace{3cm}
\end{titlepage}
\hypersetup{pageanchor=true}
\newpage

\section{Introduction}\label{Intro}
A beam of light may undergo a shift due to the occurrence of diffractive corrections, which is known as the Goos-Hänchen (GH) shift that is perpendicular to the direction of propagation in the plane containing the incident and reflected beams. The  
GH shift was discovered by
Hermann Fritz Gustav Goos and Hilda Hänchen \cite{Goos1947AP436,Goos1949AP440} and theoretically explained by Artmann using stationary phase method \cite{Artmann1948AP437} 
and Renard using energy flux method \cite{Renard1964JOSA54}. With the development of laser beam and integrated optics \cite{Lotsch1970O32-116,Lotsch1970O32-189,Lotsch1971O32-299,Lotsch1971O32-553}, the GH shift becomes very usable in the applications of optical waveguide switches 
\cite{Lotsch1970O32-189} and sensors \cite{Yin2006APL89,Tianyi2008OL33}, or in fundamental problems on tunneling time \cite{Steinberg1994PRA49,Stahlhofen2000PRA62,Balcou1997PRL78}. Recently, the lateral GH shifts have been extended to many branches of physics, such as quantum mechanics \cite{Renard1964JOSA54}, acoustics \cite{Briers2000JASA108}, neutron physics \cite{Ignatovich2004PLA322,Haan2010PRL104}, atom optics \cite{Huang2008PRA77} and spintronics \cite{Chen2008PRB77,Chen2011PRB83}. 

On the other hand, there is a big progress in studying
quantum phenomena  in graphene systems among them we cite the
quantum version of GH
shifts. 
Many
works in various graphene-based nanostructures, including single
\cite{Chen2011EPJB79}, double barrier \cite{Song2012APL100}, and superlattices
\cite{Chen2013EPJB86}, showed that the GH shifts can be enhanced by the
transmission resonances and controlled by varying the
electrostatic potential and induced gap \cite{Chen2011EPJB79}. Similar to
those in semiconductors, the GH shifts in graphene can also be
modulated by electric and magnetic
barriers \cite{Sharma2011JPCM23}. 
It has been reported that the GH shift plays an important role in
the group velocity of quasiparticles along interfaces of graphene
p-n junctions \cite{Beenakker2009PRL102,Zhao2010PRB81}.


Very recently, we have analyzed the electronic structures of massless Dirac fermions  in multi-unit graphene superlattice composed of 
3-regions \cite{kamal2018EPJB}. Subsequently,
using Chebyshev polynomials, we have studied the corresponding tunneling effect
by investigating the transmission, conductance and Fano factor \cite{Choubabi2019PSSB}.
Indeed, it was shown that
the extra Dirac points  appear at the transverse wave vector
$k_y=0$ and incident energy $\varepsilon=m \pi$ with $m$ integer value. 
In addition, at such points it was found that the transmission probability has transmission gaps, the  conductance is minimal and the Fano factor reaches a maximum. Consequently for a potential $\mathbb{V}=m\pi$ and with energy range $0\leqslant \varepsilon\leqslant\mathbb{V}$, 
 we have obtained  $(m + 1)$ transmission gaps located   at $\varepsilon=m\pi$. 

 
 Motivated by our previous works \cite{kamal2018EPJB,Choubabi2019PSSB},
 we study the GH shifts in graphene with spatially modulated potential.
 Indeed, we derive the GH shifts as function of different physical parameters based on the phase shifts in transmission and reflection. In order to provide a better insight into our results, we give a numerical study under various choices of the relevant physical parameters. In particular, we show that the GH shifts 
 can be negative as well as negative near extra Dirac points
 and can be enhanced by defect mode. Our results tell us that the GH shifts
 can be controlled 
 by 
 the width $q_2$ of the central region together with the potential height.

The present paper is organized as follows. In section \ref{Sec:Model}, for the neediness we 
review the derivation of the dispersion relation describing the energy 
of our system.
These will be used in section \ref{Sec:GH} to determine the phase shifts 
corresponding to the transmission and reflection, respectively.
Furthermore,
we derive  the GH shifts in terms of the energy, incident angle, potential height and width $q_2$.
In section \ref{Results}, we numerically analyze and discuss the transmission probability, GH shifts and the phase shifts by considering suitable choices of the physical parameters. 
We conclude our results in the final section.

\section{Theoretical model}\label{Sec:Model}

\begin{figure}[!ht]\centering
    \includegraphics[scale=0.65]{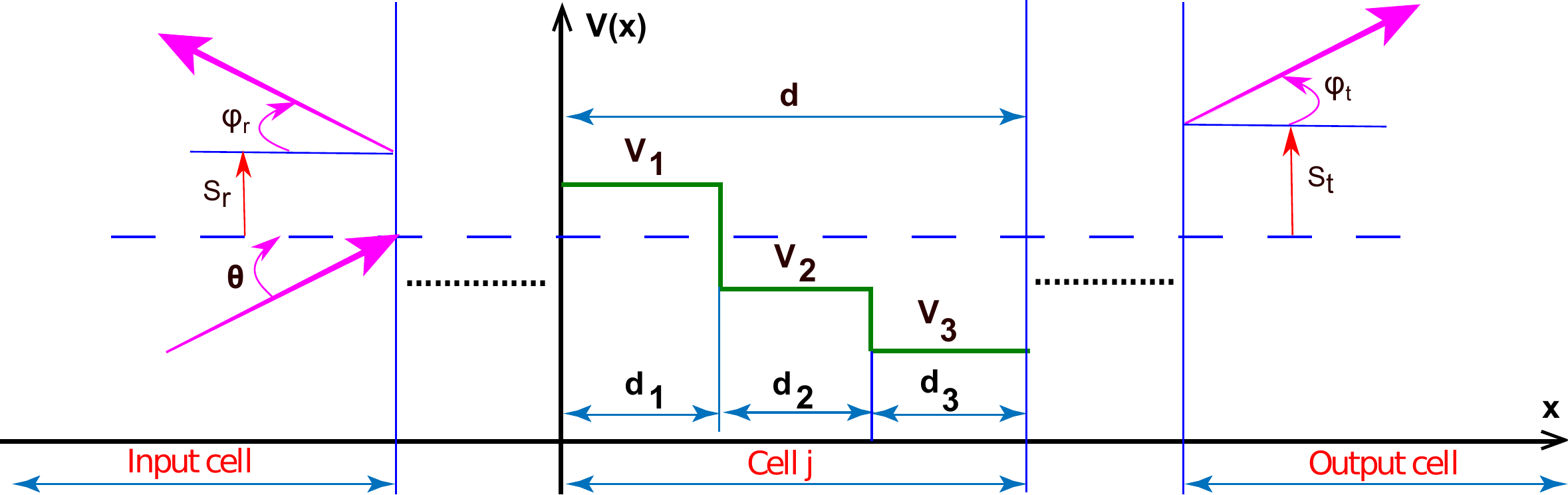}
    \caption{
        (color online) Schematic of the superlattice potential $V(x)$ composed of three regions growing along the $x$-direction. The period $d=d_1+d_2+d_3$, $d_i$ is the width of region $i$ and $V_i$ is the applied potential amplitude.
    }
    \label{Model}
\end{figure}

The Hamiltonian governing the Dirac fermions in 
graphene with spatially modulated potential can be written as
\begin{equation}
    H_{i}^{j}=v_F\bm{\sigma}\cdot\bm{p}+V_{i}^{j}\mathbb{I}
\end{equation}
where $\bm{p}=(p_x,p_y)$ is the momentum operator, $v_F\approx 10^6 \meter\per\second$ is the Fermi velocity, $\bm{\sigma}=(\sigma_x,\sigma_y)$ are the Pauli matrices, $\mathbb{I}$ is the $2\times 2$ unit matrix, the index $i$ is running from $1$ to $3$. 
Our system
is formed by $n$ unit cells that are interposed between the \textit{input} and \textit{output} regions, see Figure~\ref{Model}. The $n$ unit cells are labeled by
the index $j$ running from $0$ to $n-1$ and each $j$ is composed by a juxtaposition of the single square barriers of heights $(V_1, V_2, V_3)$ and widths $(d_1,d_2 ,d_3)$.
The potential $V(x)$ in a $j^{th }$ unit cell  is given by
\begin{equation}
    V^j(x)=
    \left\{
        \begin{array}{ccc}
            V_1,   & jd < x < d_1+jd\\
            V_2,   & d_1+jd < x < d_1+d_2+jd\\
            V_3,   & d_1+d_2+jd < x < (1+j)d.\\
        \end{array}
    \right.
\end{equation}

The Hamiltonian acts on two components of pseudospinor
$\psi_{i}(x,y) =
    \left(%
        \begin{array}{cc}
            \varphi_i^A & \varphi_i^B \\
        \end{array}%
    \right)^{T}$
where $\varphi_i^{A/B}$ are smooth enveloping functions for two triangular sublattices ($A$, $B$) in graphene and take the forms $\varphi_i^{A/B}(x) e^{\textbf{\emph{i}}k_y y}$ due to the translation invariance in the $y$-direction. The general solution is
\begin{eqnarray}
  \psi_{i}(x,y) = \psi_i(x)e^{\textbf{\emph{i}}k_y y} =
    W_i(x)A_i e^{\textbf{\emph{i}}k_y y}
\end{eqnarray}
and we have set
\begin{equation}\label{equ13}
    W_{i}(x)=
    \left(%
        \begin{array}{cc}
            e^{ \textbf{\emph{i}} k_{i} x} & e^{ -\textbf{\emph{i}} k_{i} x} \\
            s_{i} e^{ \textbf{\emph{i}} \theta_{i}} e^{ \textbf{\emph{i}} k_{i} x}  & -s_{i}
            e^{ -\textbf{\emph{i}} \theta_{i}} e^{- \textbf{\emph{i}} k_{i} x} \\
        \end{array}
    \right),\qquad A_i=
    \left(%
        \begin{array}{c}
            \alpha_i \\
            \beta_i \\
        \end{array}
    \right)
\end{equation}
with $s_i=\text{sign}(\varepsilon-\mathbb{V}_i)$, $\theta_{i} = \arctan \left(k_y/k_{i}\right)$. The parameters $ \alpha_i$, $\beta_i$ are the amplitude of positive and negative propagation wave-functions inside the region $i$, respectively. The wave vector for region $i$ takes the form
\begin{equation}\label{h0}
  k_i=\frac{1}{d}\sqrt{(\varepsilon-\mathbb{V}_i)^2-(k_yd)^2}
\end{equation}
and  the dimensionless quantities
$\mathbb{V}_i=V_i/E_F$, $\varepsilon=E/E_F$ have been introduced
with $E_F=\hbar v_F/d$, $E$ is the incident energy at angle $\theta$ of massless Dirac fermions from the \emph{input} region. At interfaces, the boundary conditions of wave-functions allow us to obtain the transfer matrix associated with $n$ identical unit cell \cite{kamal2018EPJB}
\begin{equation}\label{T17}
    \mathcal{T}_{n}=W_{0}^{-1}(0)\Omega^{n}W_{0}(nd)
\end{equation}
and $\Omega$ takes the form
\begin{equation}\label{omega}
    \Omega=W_1(0)W_1^{-1}(d_1)W_2(d_1)W_2^{-1}(d_1+d_2)W_3(d_1+d_2)W_3^{-1}(d).
\end{equation}
It is convenient to write the matrix $\Omega^{n}$ in terms of the Chebyshev polynomials of the second kind
\begin{equation}\label{362}
     \Omega^{n}= \left(%
         \begin{array}{cc}
             U_{n-1} \Omega_{11}-U_{n-2}  &  U_{n-1} \Omega_{12} \\
             U_{n-1} \Omega_{21} &  U_{n-1} \Omega_{22}-U_{n-2} \\
         \end{array}%
                \right)
\end{equation}
and
consequently, we show that 
the dispersion relation can be cast as 
\begin{equation}\label{Dispersion}
    \cos(k_x d)=\frac{1}{2}\Tr(\Omega)
\end{equation}
such that the trace is given by
\begin{eqnarray}
    \Tr(\Omega) &=& 2\left[\cos(k_1d_1)\cos(k_2d_2)\cos(k_3d_3)+Q_{12}\sin(k_1d_1)
    \sin(k_2d_2)\cos(k_3d_3)\right.\label{21}\\
    &&+\left.Q_{13}\sin(k_1d_1)\sin(k_3d_3)\cos(k_2d_2)
    +Q_{23}\sin(k_2d_2)\sin(k_3d_3)\cos(k_1d_1)\right]\nonumber
\end{eqnarray}
where we have defined the quantity
\begin{equation}
    Q_{ij}=\frac{(\mathbb{V}_i-\mathbb{V}_j)^2-(k_i^2+k_j^2)d^2}{2k_ik_jd^2},\qquad k_{i,j}\neq 0.
\end{equation}

Next we will see how the above results can be used to study the GH shifts
and investigate the basic features of
our system. This will be done by resorting in the first stage the
phase shifts associated to the transmission and reflection.

\section{Goos-Hänshen shifts}\label{Sec:GH}

To determine the GH shifts, we introduce the 
transmission and reflection probabilities, which will allow to obtain the corresponding phase shifts. Indeed,
knowing that the amplitudes $A_{in}=
    \left(%
        \begin{array}{c}
            1 \\
            r_n \\
        \end{array}%
    \right)$ and $ A_{out}=
    \left(%
        \begin{array}{c}
            t_n \\
            0 \\
        \end{array}%
    \right)$
of the eigenspinors in {\it input} and {\it output} regions are connected by the  relation
\begin{equation}\label{35}
    A_{in}=\mathcal{T}_{n} A_{out}
\end{equation}
we show that
the solution of \eqref{35} provides the transmission and reflection amplitudes ($t_n$, $r_n$) in terms of transfer matrix elements, which is
\begin{equation}\label{361B}
    t_{n}=\frac{1}{\mathcal{T}_{n_{11}}},\qquad r_{n}=\frac{\mathcal{T}_{n_{21}}}{\mathcal{T}_{n_{11}}}.
\end{equation}
To proceed further, we use 
the complex notation to write
\begin{equation}\label{trref}
 t_n = \rho_t e^{\textbf{\emph{i}}\varphi_{t}},
\qquad r_n = \rho_r e^{\textbf{\emph{i}}\varphi_{r}}
\end{equation}
such that the   phase shifts and modulus
are given by
\begin{eqnarray}
  &&  \varphi_{t} = \arctan\left(\textbf{\emph{i}}\frac{t_n^{\ast}-t_n}{t_n+t_n^{\ast}}\right),\qquad \varphi_{r} =\arctan\left(\textbf{\emph{i}}\frac{r_n^{\ast}-r_n}{r_n+r_n^{\ast}}\right)\\
&&
    \rho_{t} =\sqrt{\Re^2 [t_n]+\Im^2 [t_n]}, \qquad \rho_{r} =\sqrt{\Re^2 [r_n])+\Im^2 [r_n]}.
\end{eqnarray}
From the above results, we can easily derive the corresponding transmission $T_n$ and reflection $R_n$ probabilities. They are
\begin{equation}
    T_n=\rho_{t}^{2},\qquad  R_n=\rho_{r}^{2}.
\end{equation}

To study the  GH shifts in graphene with spatially modulated potential, we consider the incident, reflected and transmitted beams around some transverse wave vector $k_{y_0}$ corresponding to the incidence angle $\theta_{0}(k_{y_{0}})$, which belongs to the interval $[0, \pi/2]$. Indeed, the incident beam is given by
\begin{eqnarray}
   \Psi_{in}(x,y) &=& \dfrac{1}{\sqrt{2\pi}}\int_{-\infty}^{+\infty}dk_y\ f(k_y-k_{y_0})\ e^{\textbf{\emph{i}}(k_{in}x+k_yy)}\left(
            \begin{array}{c}
              {1} \\
              {e^{\textbf{\emph{i}}\theta(k_{y})}}
            \end{array}
          \right)\label{eq79}
\end{eqnarray}
where $k_{in}$ is the $x$-component of wave vector in the \emph{input} region, $k_{in}=\sqrt{\varepsilon^2-(k_yd)^2}/d$, $\theta = \arctan \left(k_y/k_{in}\right)$ by assuming $\mathbb{V}_{in}=0$. The reflected beam takes the form
\begin{eqnarray}
\Psi_{re}(x,y) &=& \dfrac{1}{\sqrt{2\pi}}\int_{-\infty}^{+\infty}dk_y\ r_n(k_y)\
f(k_y-k_{y_0})\ e^{\textbf{\emph{i}}(-k_{in}(k_y)x+k_yy)}\left(
            \begin{array}{c}
              {1} \\
              {-e^{-\textbf{\emph{i}}\theta(k_{y})}} \\
            \end{array}
          \right)\label{refl}
\end{eqnarray}
and $f(k_y-k_{y_0}) =w_ye^{-w_{y}^2(k_y-k_{y_0})^2/2}$ is the angular spectral distribution, assumed of Gaussian shape,
 $w_y=w\sec\theta_0$ and $w$ is the half beam width at waist \cite{Beenakker2009PRL102}. When the electron beam is well-collimated, there will be a narrow distribution of $k_y$ values around $k_{y_0}$corresponding to a small beam divergence $\delta\theta=\lambda/\pi w$ with $\lambda=2\pi d/\varepsilon$ is the Fermi length. With this assumption, $k_{y}$-dependent terms can be approximated by a Taylor expansion around ${k_y}_0$  retaining only the first order term we get
\begin{eqnarray}
&&\theta(k_{y})\approx
\theta(k_{y_{0}})+(k_{y}-k_{y_{0}})\eval{\frac{\partial\theta}{\partial
k_{y}}}_{k_{y_{0}}}
\\
&&
k_{in}(k_{y})\approx k_{in}(k_{y_{0}})+(k_{y}-k_{y_{0}})\eval{\frac{\partial
k_{in}}{\partial k_{y}}}_{k_{y_{0}}}.
\end{eqnarray}
Evaluating the Gaussian integral \eqref{eq79} 
to obtain the  the incident beam  $\Psi_{inc}^\pm=e^{-(y-{y_0}_{inc}^\pm)^2/2w_y^2}$ at $x=0$ such that 
the two components are 
\begin{equation}
   {y_0}_{inc}^+=0,\qquad {y_0}_{inc}^-=-\eval{\frac{\partial\theta}{\partial
k_{y}}}_{k_{y_{0}}}
\end{equation}
giving rise to the separation of the two centers of the incident beam
\begin{equation}
   \delta_0=|{y_0}_{inc}^+-{y_0}_{inc}^-|=-\eval{\frac{\partial\theta}{\partial
k_{y}}}_{k_{y_{0}}}.
\end{equation}
For convenience, we assume that the potential is the same in the \emph{input} and \emph{output} region, then
the transmitted beam can be  expressed as
\begin{eqnarray}
\Psi_{tr}(x,y) &=& \dfrac{1}{\sqrt{2\pi}}\int_{-\infty}^{+\infty}dk_y\ t_n(k_y)\
f(k_y-k_{y_0})\ e^{\textbf{\emph{i}}[k_{in}(k_y)(x-L)+k_yy]}\left(
            \begin{array}{c}
              {1} \\
              {e^{\textbf{\emph{i}}\theta(k_{y})}} \\
            \end{array}
          \right)\label{trans}
\end{eqnarray}
where $L$ is the total length of the graphene superlattice. The stationary phase approximation requires that the phase $\varphi_t$ of the transmission coefficient $t_n(k_y)$ is linearly dependent on $k_y$ and the amplitude of the transmission keeps almost constant, allowing to expand  $\varphi_t(k_y)$ in Taylor series at ${k_y}_0$, and retain up to the first-order term. After substituting all into the Gaussian integral \eqref{trans}, we obtain $\Psi_{tr}^\pm=e^{-(y-{y_0}_{tr}^\pm)^2/2w_y^2}$ at $x=L$ with the two components of the transmitted beam 
\begin{equation}
    {y_0}_{tr}^+=-\eval{\frac{\partial\varphi_t}{\partial
k_{y}}}_{k_{y_{0}}},\qquad {y_0}_{tr}^-=-\eval{\frac{\partial\varphi_t}{\partial
k_{y}}}_{k_{y_{0}}}-\eval{\frac{\partial\theta}{\partial
k_{y}}}_{k_{y_{0}}}.
\end{equation}
We notice that the separation $\delta_0$ of the two centers for the transmitted beam is the same as that for the incident beam. In addition, the comparison of the incident and transmitted beams, leads to the displacement of the two components
\begin{equation}
   s_t^\pm=|{y_0}_{tr}^\pm-{y_0}_{inc}^\pm|=-\eval{\frac{\partial\varphi_t}{\partial
k_{y}}}_{k_{y_{0}}}
\end{equation}
which can be used to define 
 the GH shifts in transmission 
 \cite{Beenakker2009PRL102}
\begin{equation}
   S_{t}=\dfrac{s_t^++s_t^-}{2}.
\end{equation}
In similar way we can derive the GH shifts in the reflected electron beam in graphene 
with spatially modulated potential
using the reflection coefficient obtained before. Consequently, 
the two GH shifts are given by
\begin{equation}
        S_{t}=- \frac{\partial \varphi_{t}}{\partial
        k_{y}}\Big|_{k_{y0}}, \qquad S_{r}=- \frac{\partial \varphi_{r}}{\partial
        k_{y}}\Big|_{k_{y0}}.
\end{equation}
In the next we focus only the numerical study of the GH shifts in transmission
and the other one can done in similar way. 

\section{Numerical results}\label{Results}

We will numerically analyze and discuss the the properties of GH shifts near 
the extra Dirac points located at finite energy $\varepsilon=m\pi$, with $m$  integer, for Dirac fermions in graphene with spatially modulated potential. This will be done by tuning on different physical parameters characterizing our system under various conditions.
\begin{figure}[!hbt]\centering
    \subfloat[$q_2=0$]{\centering
        \includegraphics[width=3.9cm]{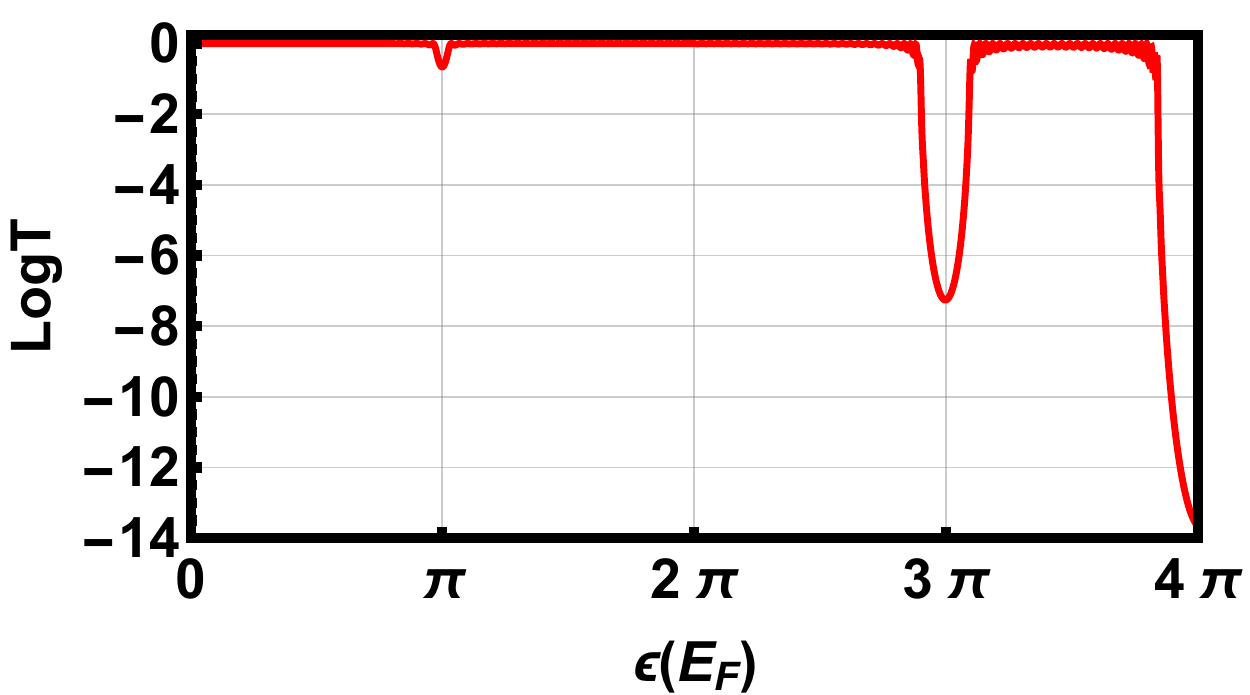}
        \label{FigLnTq2:A}
    }\hspace{-0.25cm}
    \subfloat[$q_2=1/4$]{\centering
        \includegraphics[width=3.95cm]{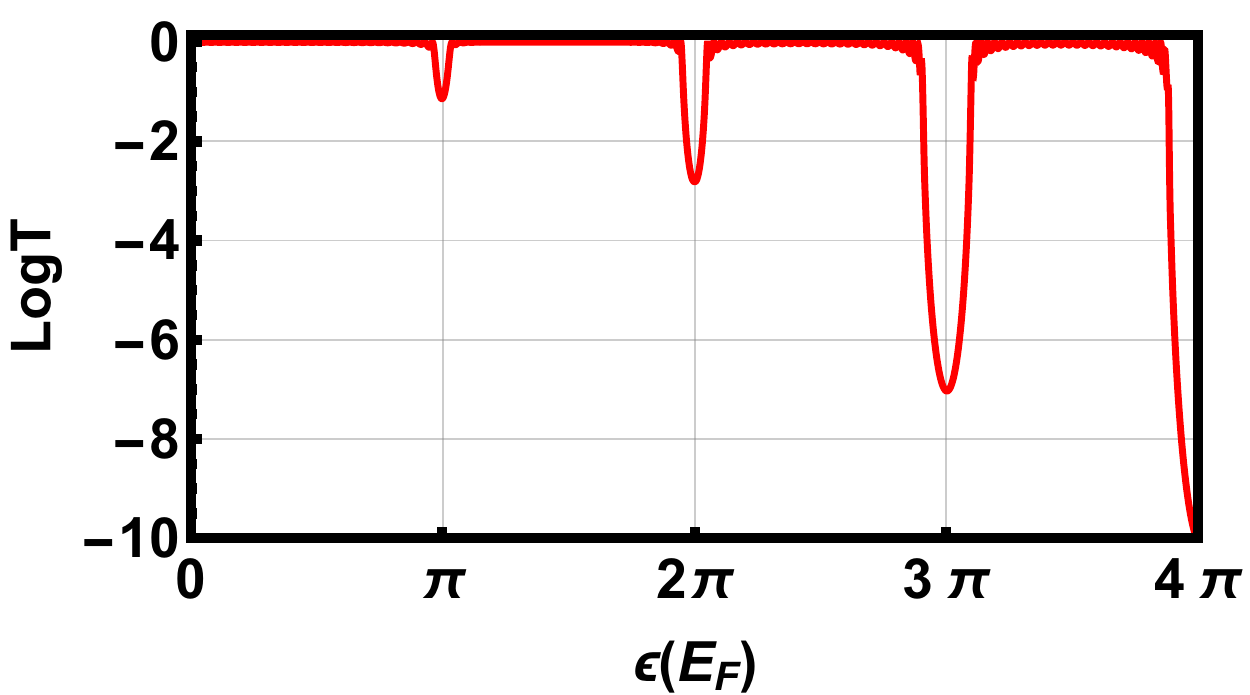}
        \label{FigLnTq2:B}
    }\hspace{-0.25cm}
    \subfloat[$q_2=1/2$]{\centering
        \includegraphics[width=3.95cm]{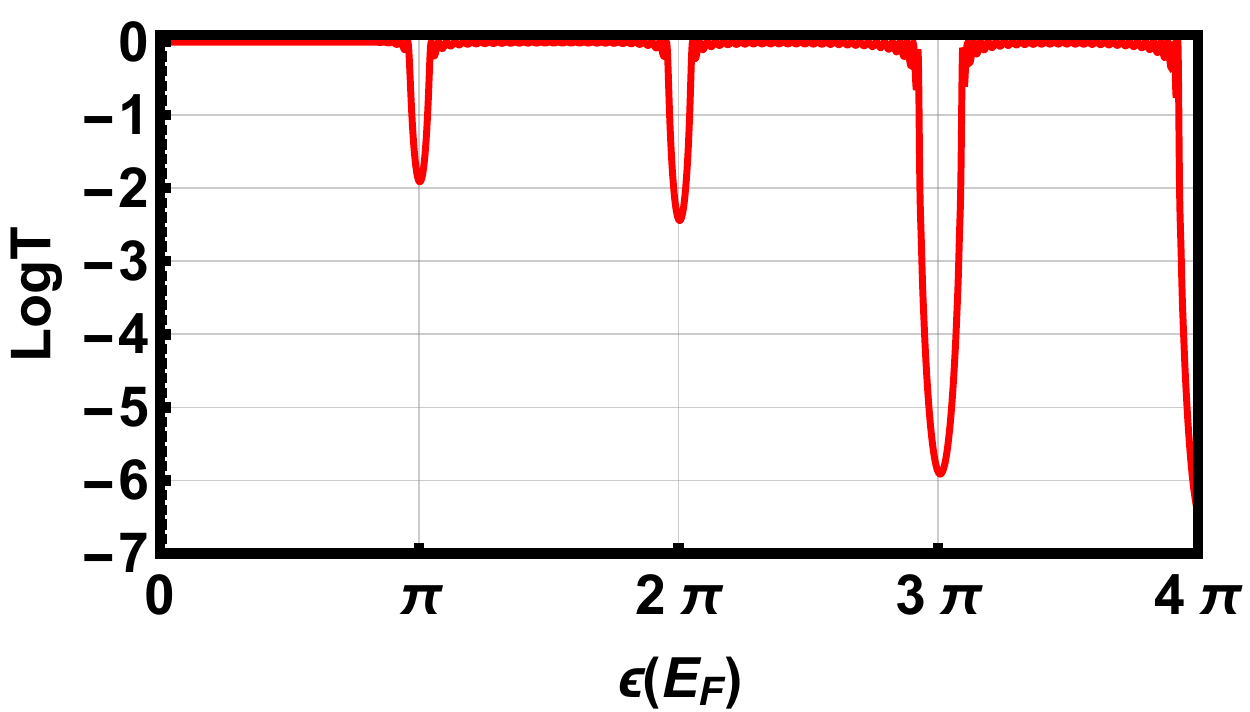}
        \label{FigLnTq2:C}
    }\hspace{-0.25cm}
    \subfloat[$q_2=3/4$]{\centering
        \includegraphics[width=3.95cm]{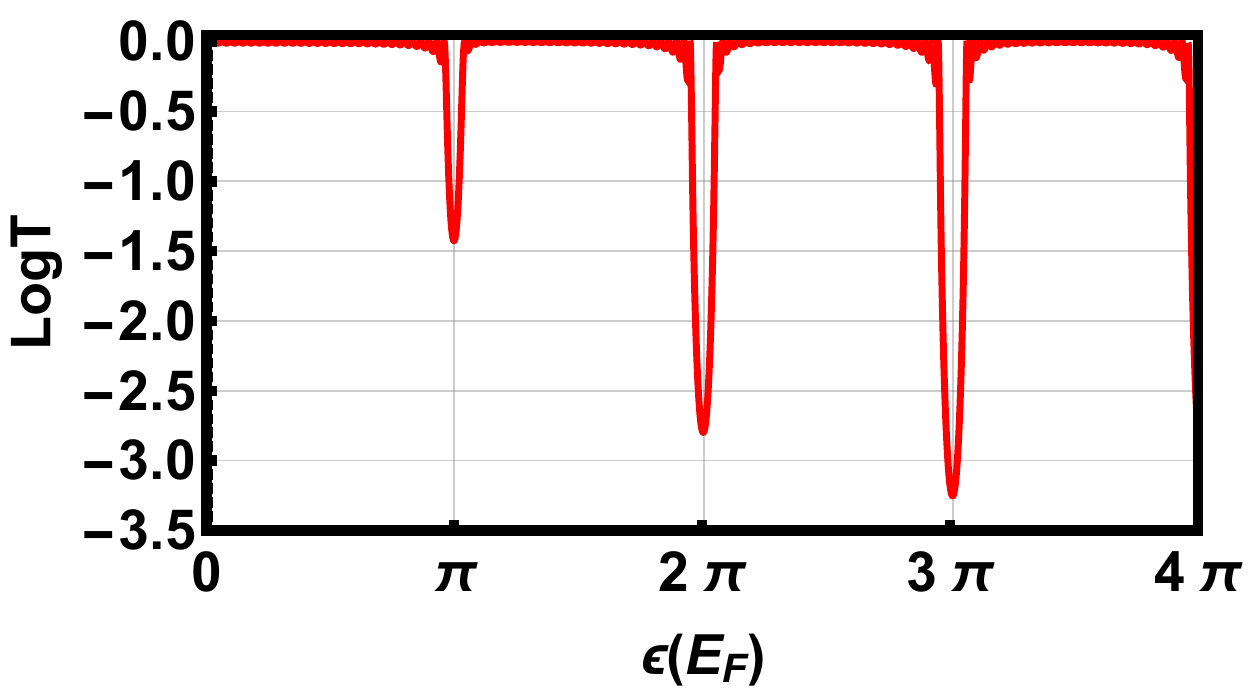}
        \label{FigLnTq2:D}
    }\vspace{-0.5cm}
    \subfloat[$q_2=0$]{\centering
        \includegraphics[width=3.95cm]{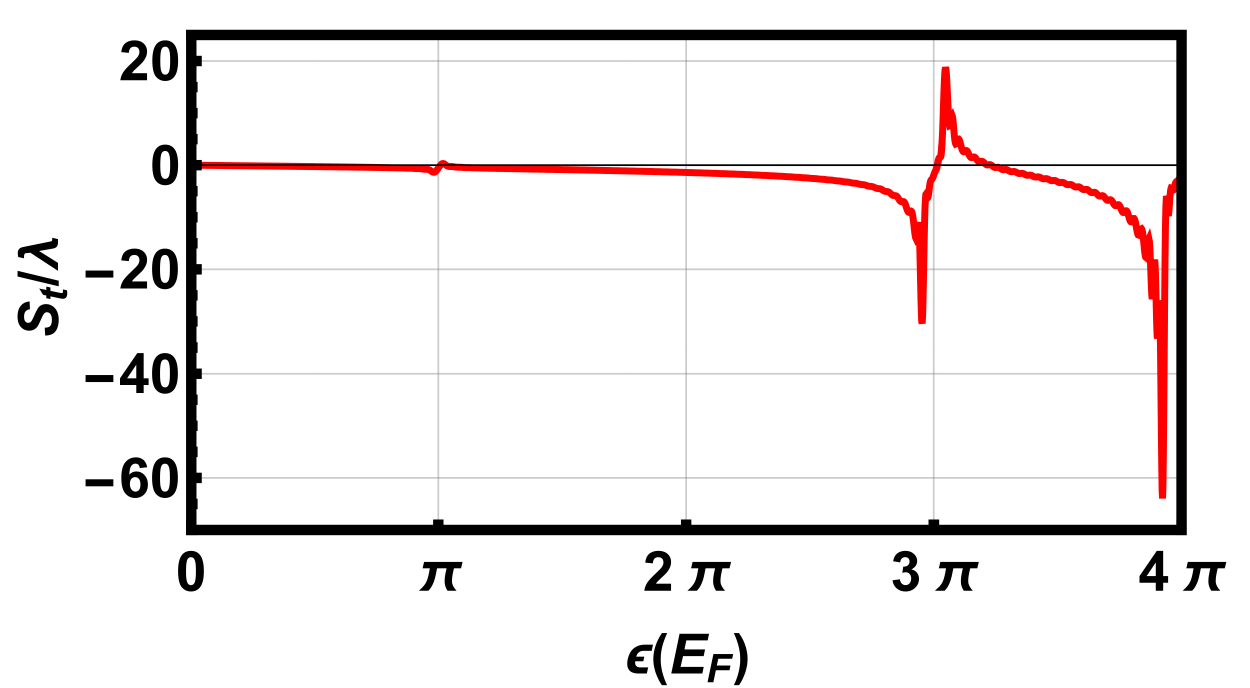}
        \label{FigStq2:A}
    }\hspace{-0.25cm}
    \subfloat[$q_2=1/4$]{\centering
        \includegraphics[width=3.95cm]{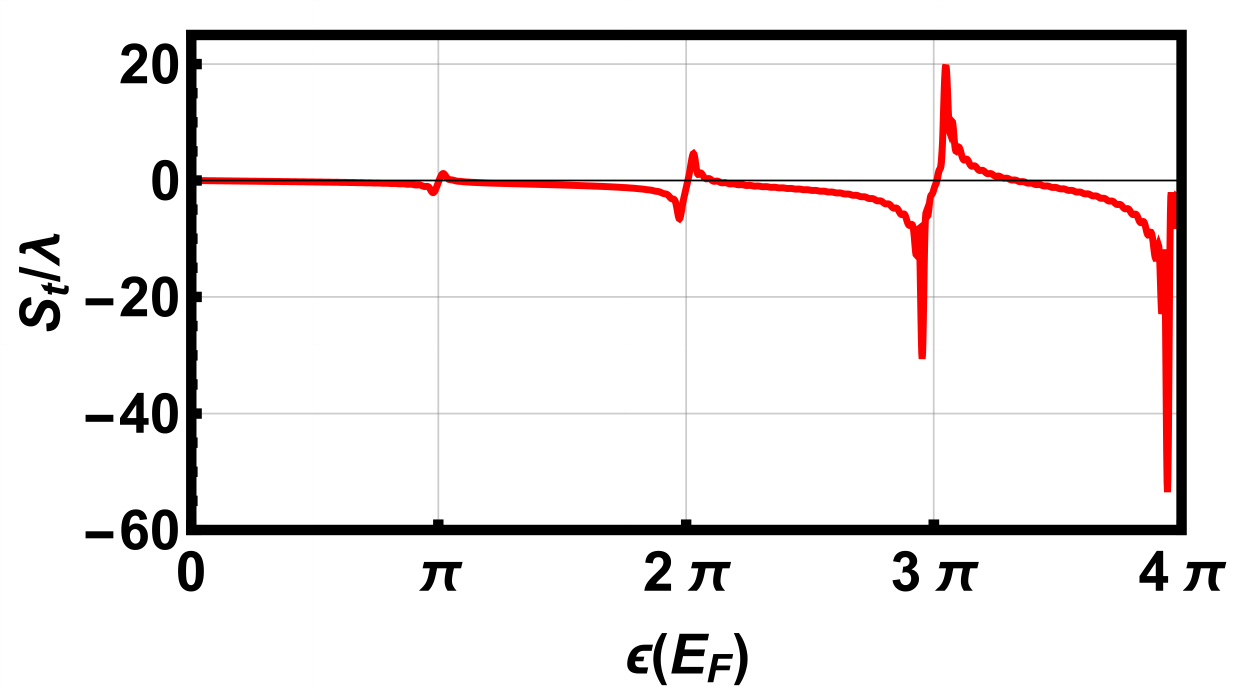}
        \label{FigStq2:B}
    }\hspace{-0.25cm}
    \subfloat[$q_2=1/2$]{\centering
        \includegraphics[width=3.95cm]{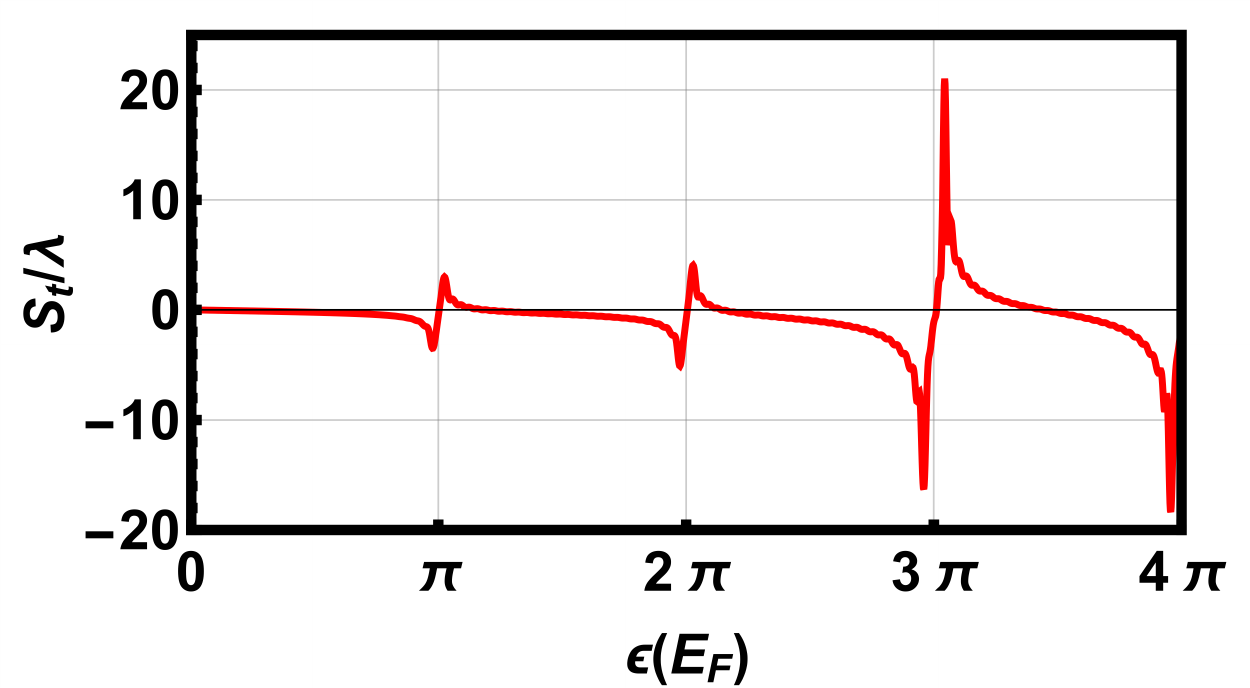}
        \label{FigStq2:C}
    }\hspace{-0.25cm}
    \subfloat[$q_2=3/4$]{\centering
        \includegraphics[width=3.95cm]{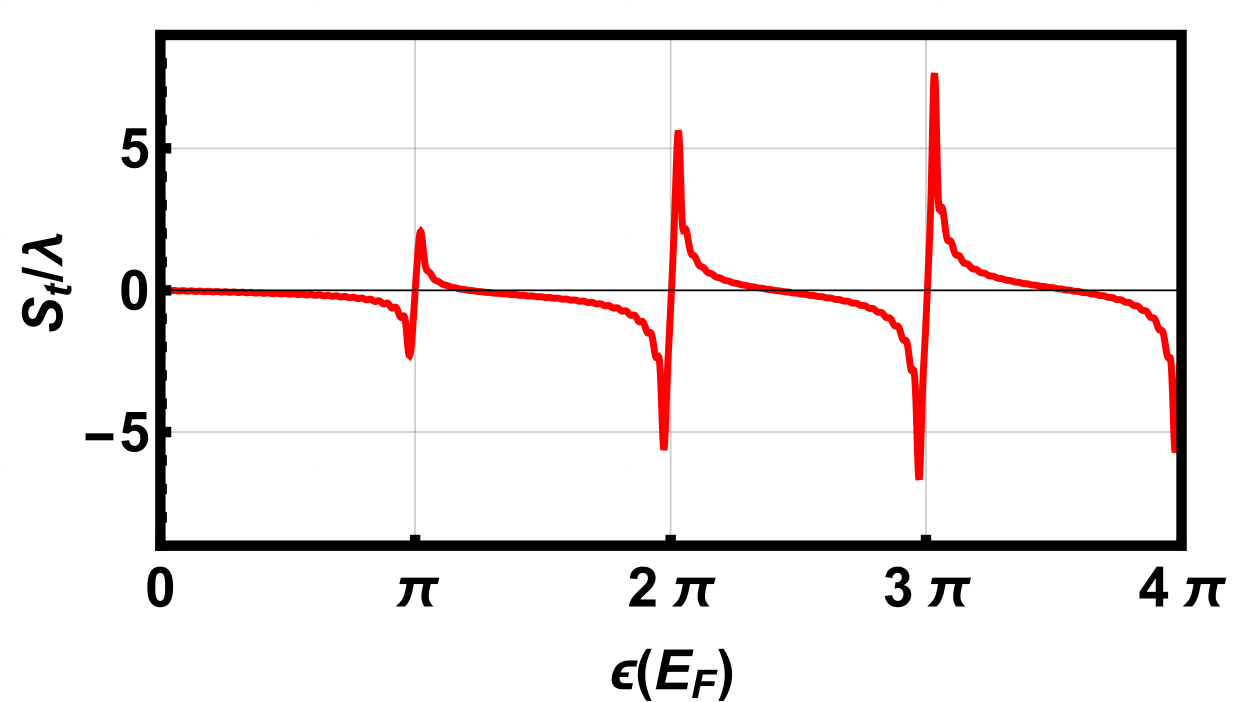}
        \label{FigStq2:D}
    }
     \caption{
        (color online) \protect\subref{FigLnTq2:A}-\protect\subref{FigLnTq2:D}: transmission probability and \protect\subref{FigStq2:A}-\protect\subref{FigStq2:D}: the  GH shifts $S_t/\lambda$ versus the incident energy $\varepsilon$ for the transmitted electron beam in the graphene with spatially modulated potential of $n=30$ unit cells, for $\theta_0=5\degree$, $\mathbb{V}=4\pi$ and four values of the width $q_2=0, \frac{1}{4}, \frac{1}{2}, \frac{3}{4}$.
     }
    \label{FigLnTq2}
\end{figure}
Indeed, Figure \ref{FigLnTq2} shows the transmission probability (Figures \subref{FigLnTq2:A}-\protect\subref{FigLnTq2:D}) and the corresponding GH shifts $S_t/\lambda$ (Figures \subref{FigStq2:A}-\subref{FigStq2:D}) versus the incident energy $\varepsilon$ for the transmitted electron beam 
with the configuration of the parameters $n=30$, $\theta_0=5\degree$, $\mathbb{V}=4\pi$ for different values of $q_2$. We observe that  whatever the value of the distance $q_2$, the extra Dirac points \cite{kamal2018EPJB} and transmission gaps \cite{Choubabi2019PSSB}, as shown in Figure \ref{FigLnTq2}(\subref{FigLnTq2:A}-\protect\subref{FigLnTq2:D}), are located at finite energy $\varepsilon_D=m\pi$, $m$ integer. Obviously, near the extra Dirac points at finite energy $\varepsilon_D$ and for any distance $q_2$ the GH shifts are always negative when $\varepsilon<\varepsilon_D$ and  
positive in the opposite side, namely $\varepsilon>\varepsilon_D$. In addition,
we notice that the negative and positive GH shifts, associated with extra Dirac points at finite energy, are sensitive to the parameter $q_2$. In fact, as long as $q_2$ increases the GH shifts decrease, which may be useful to control their negative and positive behaviors. 

\begin{figure}[!hbt]\centering
    \subfloat[]{\centering
        \includegraphics[height=4.25cm]{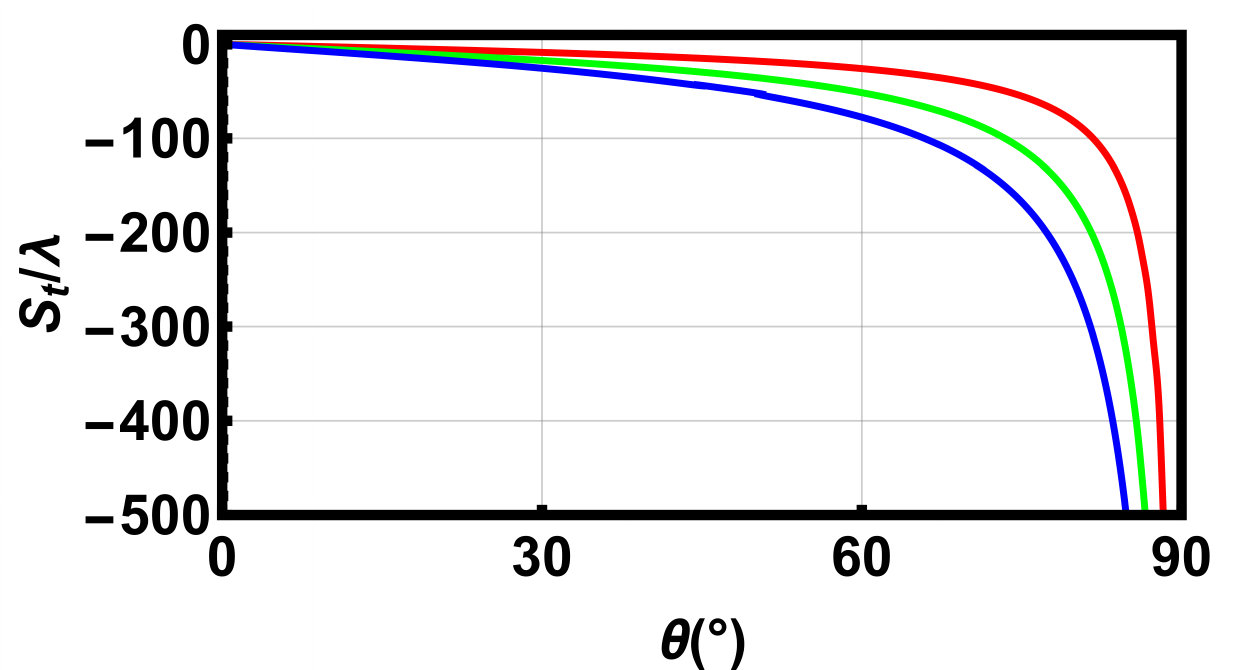}
        \label{FigStETheta:A}
    }
    \subfloat[]{\centering
         \includegraphics[height=4.25cm]{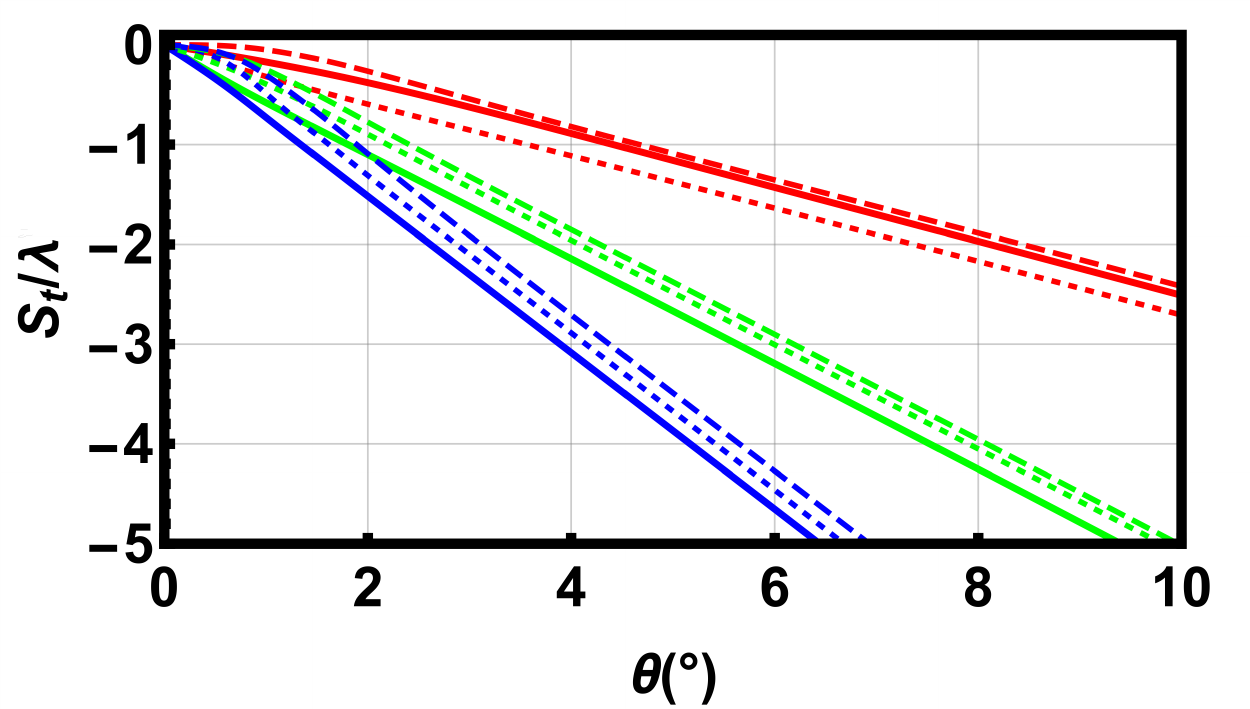}
        \label{FigStETheta:B}
    }
     \caption{
     (color online) \protect\subref{FigStETheta:A}: The GH shifts $S_t/\lambda$ versus the incident angle $\theta$ for the transmitted electron beam in graphene with spatially modulated potential of $n=30$ unit cells, for   $q_2=1/3$, $\mathbb{V}=4\pi$ and three values of incident energy $\varepsilon=\pi$ (red line), $2\pi$ (green line), $3\pi$ (blue line). \protect\subref{FigStETheta:B}: $S_t/\lambda$ versus small $\theta$
     with the same conditions but for three values of width $q_2=1/3$ (thick line), $1/2$ (dotted line), $2/3$ (dashed line).
     }
    \label{FigStETheta}
\end{figure}

Figure \ref{FigStETheta} presents the behaviors of 
the GH shifts $S_t/\lambda$ versus the incident angle $\theta$ for the transmitted electron beam in graphene with spatially modulated potential of $n=30$ unit cells. Indeed, 
in Figure \ref{FigStETheta}\protect\subref{FigStETheta:A} 
we chose the parameter configuration $q_2=1/3$, $\mathbb{V}=4\pi$ and three values of the incident energy $\varepsilon=\pi$ (red line), $2\pi$ (green line), $3\pi$
(blue line). We observe that at normal incidence, i.e. $\theta=0$, the  GH  shifts  at extra Dirac points are null, which is the result of the pristine graphene. However, when $\theta$ increases $S_t/\lambda$ also increase and for $\theta=90\degree$  become giant. On the other hand, it is clearly seen that when the quantum number $m$ increases the GH shifts increase slowly. {In order to characterize the GH shift behaviors for small values 
$\theta$ and different values of width, we present in Figure \ref{FigStETheta}\protect\subref{FigStETheta:B} $S_t/\lambda$ versus  $\theta$ for three values 
$\varepsilon=\pi$, $2\pi$, $3\pi$ highlighted by red, green and blue lines, respectively and for three values 
$q_2=1/3$ (thick line), $1/2$ (dotted line), $2/3$ (dashed line).
It is interesting to note that for $\varepsilon=2\pi$, $3\pi$ as long as $q_2$ increases $S_t/\lambda$ increase, but for $\varepsilon=2\pi$ the GH shifts corresponding to $q_2=2/3$ is lower than that corresponding to $q_2=1/3$}. 
This result shows the influence of the width $q_2$ on the behavior of $S_t/\lambda$
and then tells us that 
one may  manipulate 
the GH shifts by playing on $q_2$. This statement will be clarified in the next Figure.


\begin{figure}[!hbt]\centering
    \subfloat[$\theta=5\degree$]{\centering
        \includegraphics[width=5.25cm]{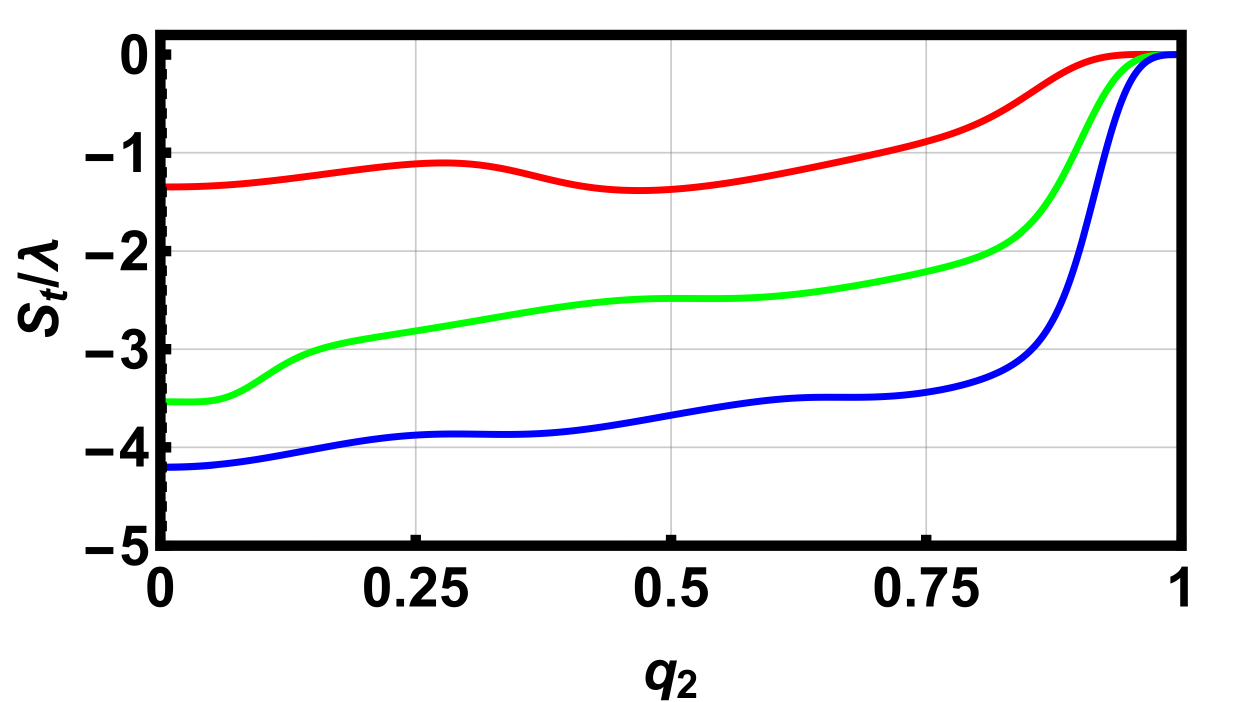}
        \label{}
    }
    \subfloat[$\theta=10\degree$]{\centering
        \includegraphics[width=5.25cm]{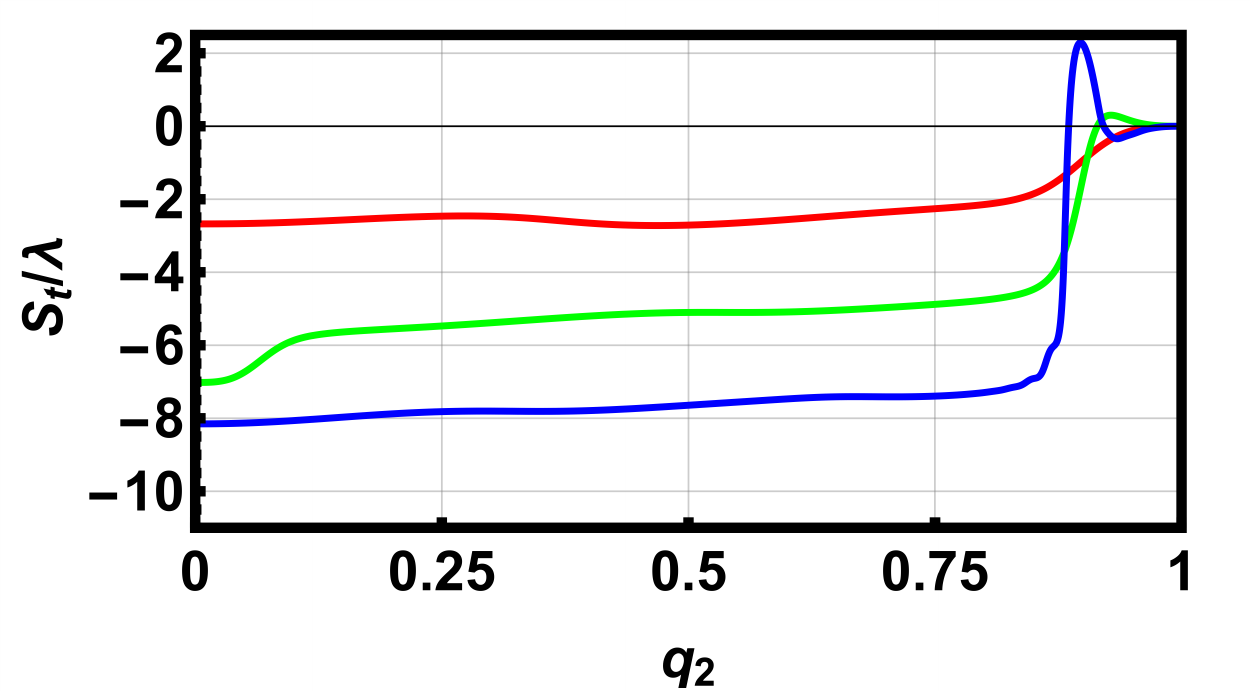}
        \label{}
    }
    \subfloat[$\theta=15\degree$]{\centering
        \includegraphics[width=5.25cm]{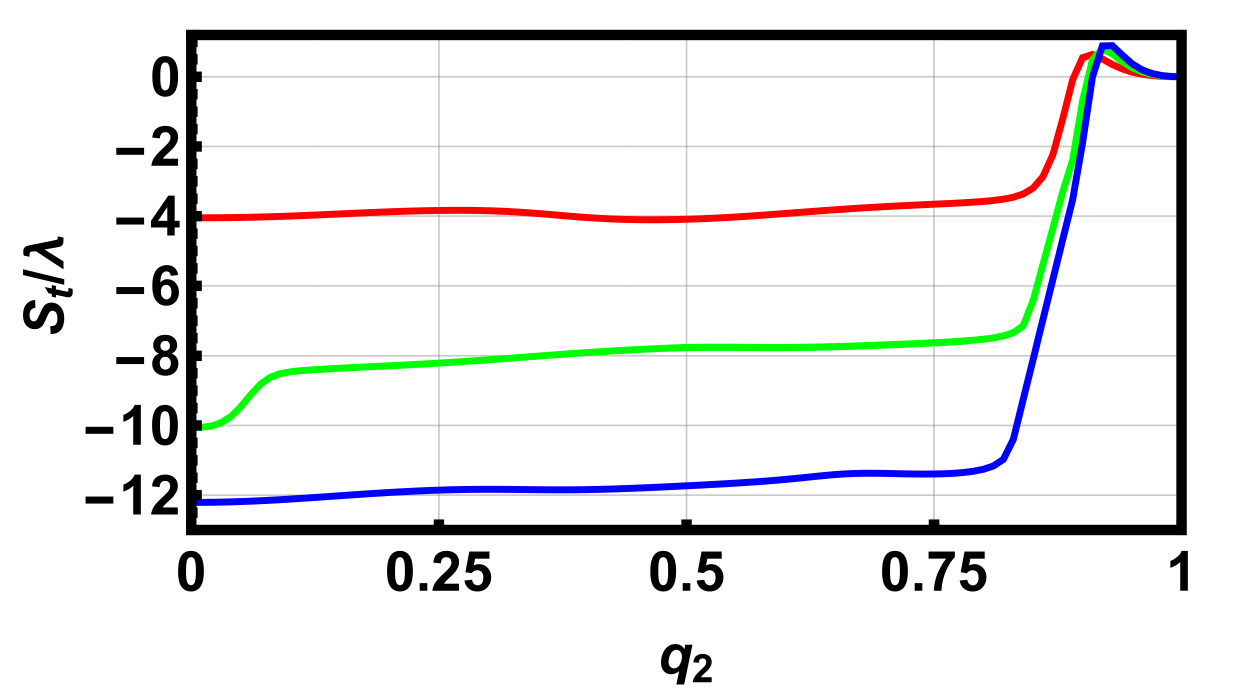}
        \label{}
    }
     \caption{
     (color online) The GH shifts $S_t/\lambda$ versus the distance $q_2$ for the transmitted electron beam in graphene with spatially modulated potential of $n=30$ unit cells, for $\mathbb{V}=4\pi$, $\theta=5\degree$ and four values of incident energy $\varepsilon=\pi$ (red line), $2\pi$ (green line), $3\pi$ (blue line), $4\pi$ (magenta line).
     }
    \label{FigStq2E}
\end{figure}

Figure \ref{FigStq2E} illustrates the GH  shifts $S_t/\lambda$ versus the distance $q_2$ for the transmitted electron beam in graphene with spatially modulated potential of $n=30$ unit cells, for
$\mathbb{V}=4\pi$, $\theta=5\degree$ and four values of incident energy $\varepsilon=\pi$ (red line), $2\pi$ (green line), $3\pi$ (blue line), $4\pi$ (magenta line).
%
One can see that whatever $q_2$, for $\theta=5\degree$ and $\varepsilon=m\pi$, as long as $m$ increases $S_t/\lambda$ decrease negatively. In addition, for $q_2$ between $0$ and $0.75$ the GH  shifts vary slightly, but from the value $q_2=0.75$ it  increase rapidly until  become null for $q_2=1$ that is the case of the pristine graphene. For $q_2$ near $1$, when $\theta$ increases $S_t/\lambda$ change sign from negative to positive and then go to zero for $q_2=1$. 
One remarkably effect to mention is that when $\theta=10\degree$, we observe the GH shifts are almost constant till some values near $q_2=1$. This tell us that the variation of  $S_t/\lambda$ in terms of the width $q_2$ is strongly affected by the values
taken by the incident angle $\theta$ and therefore it can be used to act on 
$S_t/\lambda$.


\begin{figure}[!hbt]\centering
    \subfloat[$q_2=1/3$]{\centering
        \includegraphics[width=5.25cm]{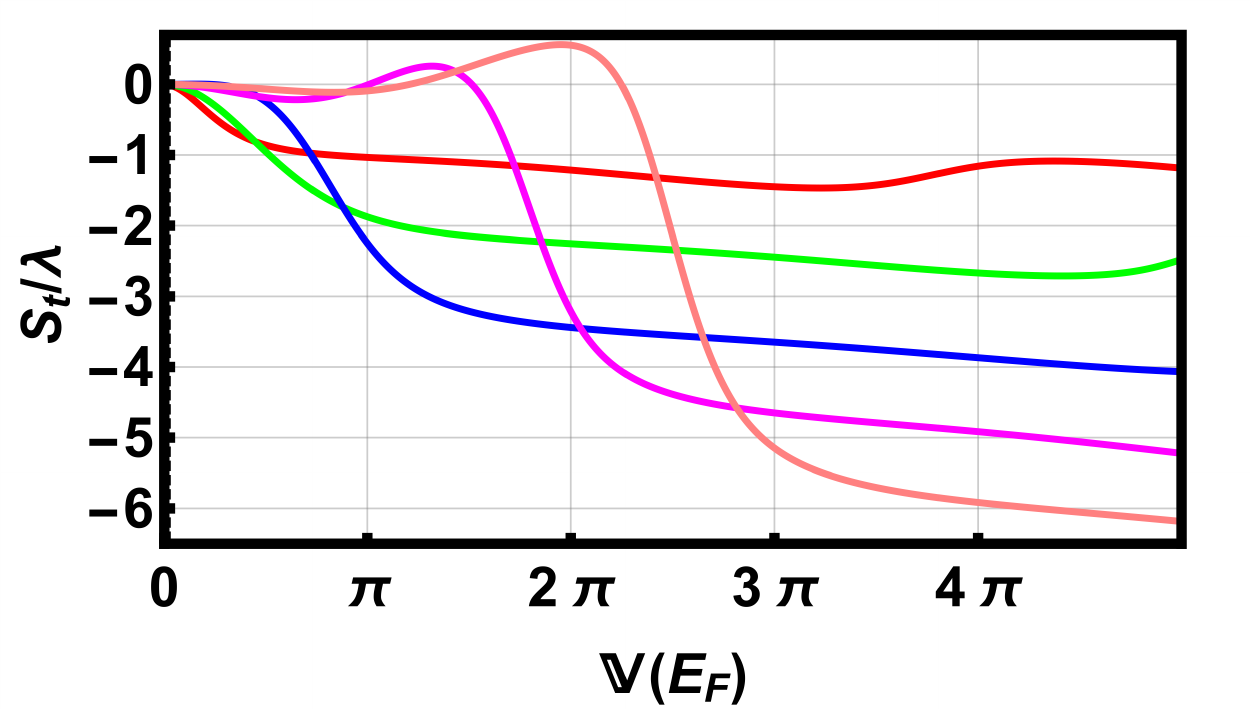}
        \label{StVqETheta5:A}
    }
    \subfloat[$q_2=1/2$]{\centering
        \includegraphics[width=5.25cm]{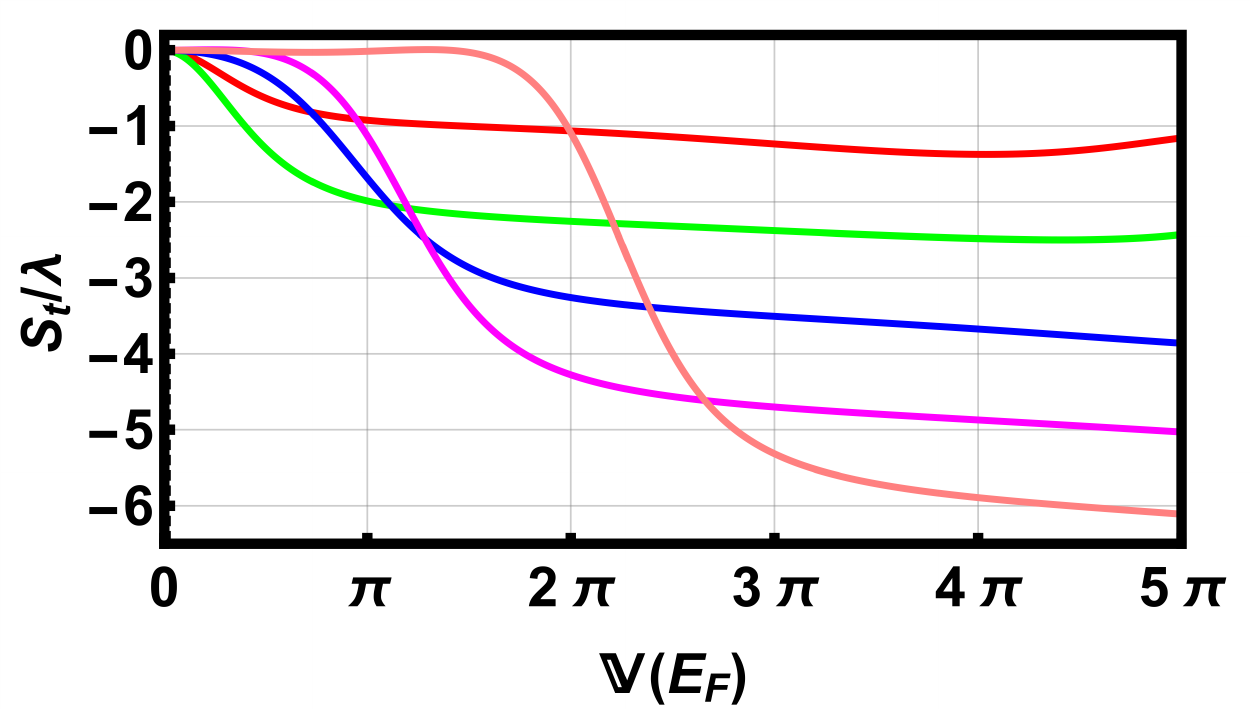}
        \label{StVqETheta5:B}
    }
    \subfloat[$q_2=2/3$]{\centering
        \includegraphics[width=5.25cm]{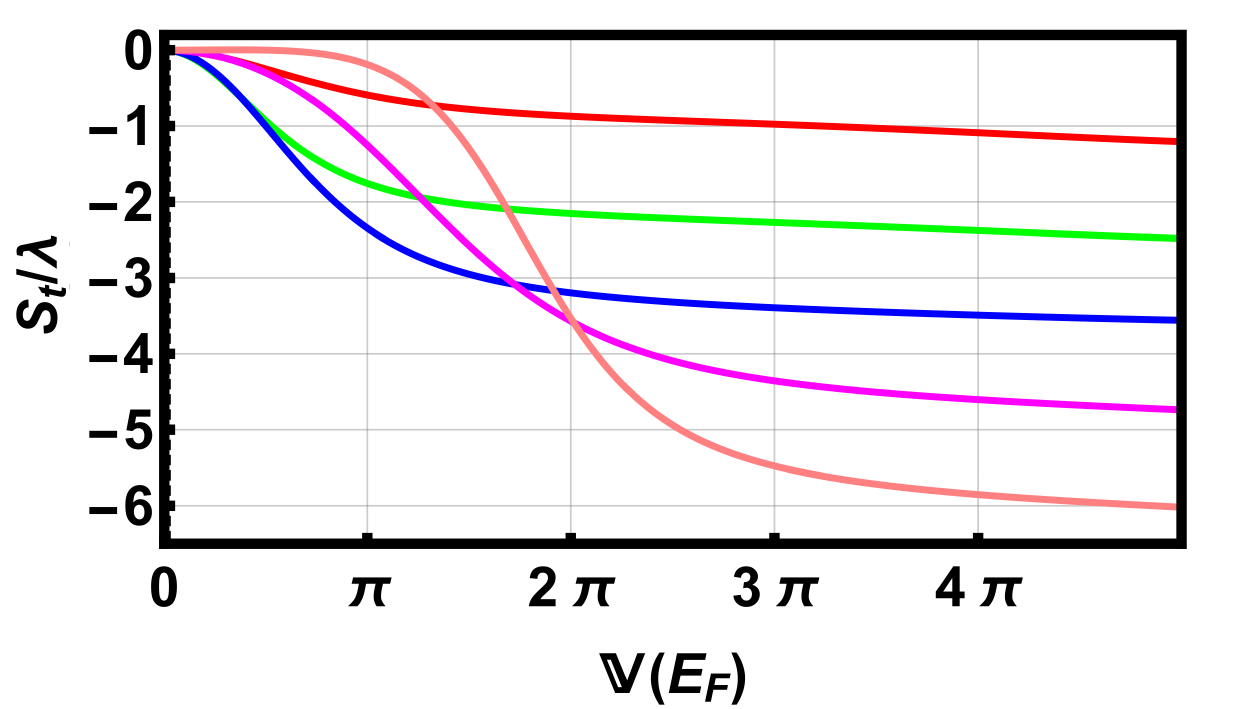}
        \label{StVqETheta5:C}
    }
     \caption{
     (color online) The GH shifts $S_t/\lambda$ versus the potential $\mathbb{V}$ for the transmitted electron beam in graphene with spatially modulated potential of $n=30$ unit cells, for $\theta=5\degree$ for five values of incident energy $\varepsilon=\pi$ (red line), $2\pi$ (green line), $3\pi$ (blue line), $4\pi$ (magenta line), $5\pi$ (pink line).
     \protect\subref{StVqETheta5:A}: $q_2=1/3$, \protect\subref{StVqETheta5:B}: $q_2=1/2$, \protect\subref{StVqETheta5:C}: $q_2=2/3$.
     }
    \label{FigStVq13E}
\end{figure}

We present in Figure \ref{FigStVq13E} the GH shifts $S_t/\lambda$ versus the potential $\mathbb{V}$ for the transmitted electron beam in graphene with spatially modulated potential of $n=30$ unit cells, for $q_2=1/3$, $\theta=5\degree$ and four values of incident energy $\varepsilon=\pi$ (red line), $2\pi$ (green line), $3\pi$ (blue line), $4\pi$ (magenta line), $5\pi$ (pink line). We notice that, for  $\varepsilon=\pi$, $2\pi$, $3\pi$ and $q_2=1/3$ the GH  shifts are always negative and decrease when the potential increases. For $\varepsilon=4\pi$, $5\pi$, as long as the potential increases, $S_t/\lambda$ start with negative value and increase to reach a positive peak, then decrease rapidly to negative regime. According to Figures \ref{FigStVq13E}\subref{StVqETheta5:B}, \ref{FigStVq13E}\subref{StVqETheta5:C} we observe that  when $q_2$ increases, $S_t/\lambda$ decrease rapidly to negative regime. Therefore, the GH shifts $S_t/\lambda$ can be changed from positive to negative by controlling the strength of the potential $\mathbb{V}$ and the width $q_2$.

\begin{figure}[!hbt]\centering
    \subfloat[]{\centering
        \includegraphics[width=5.25cm]{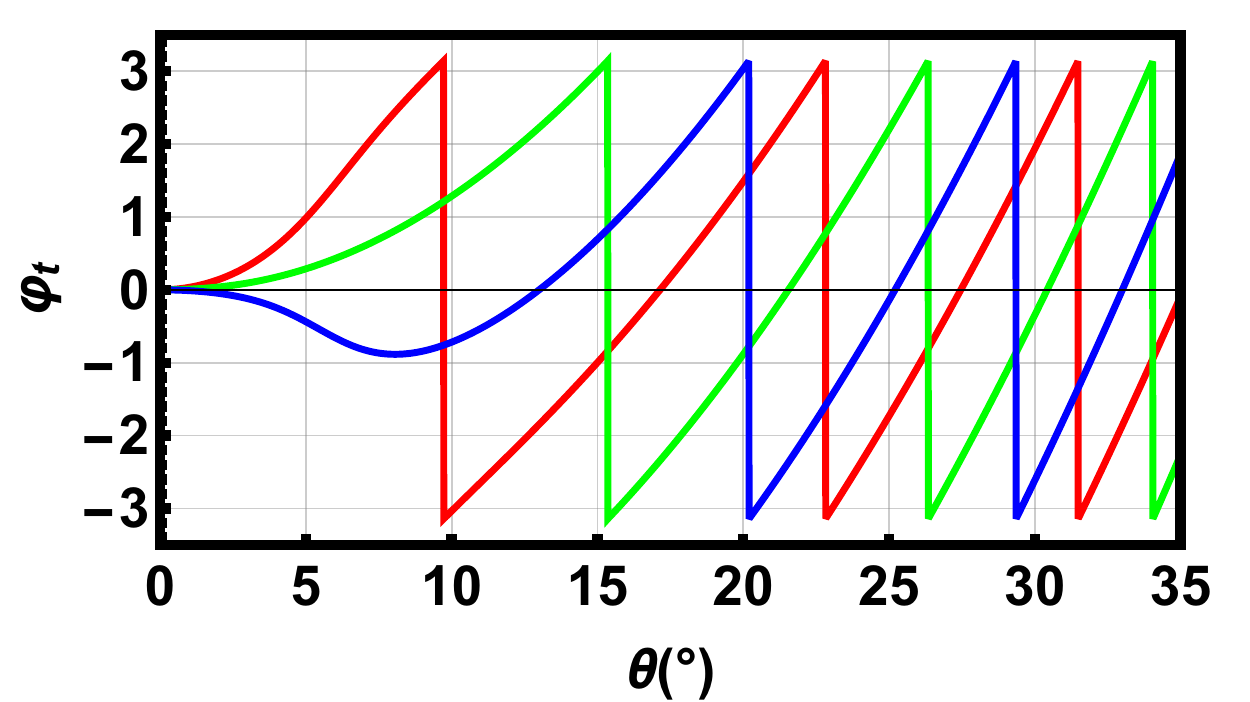}
        \label{FigPhiThetaE1pi:A}
    }
    \subfloat[]{\centering
        \includegraphics[width=5.25cm]{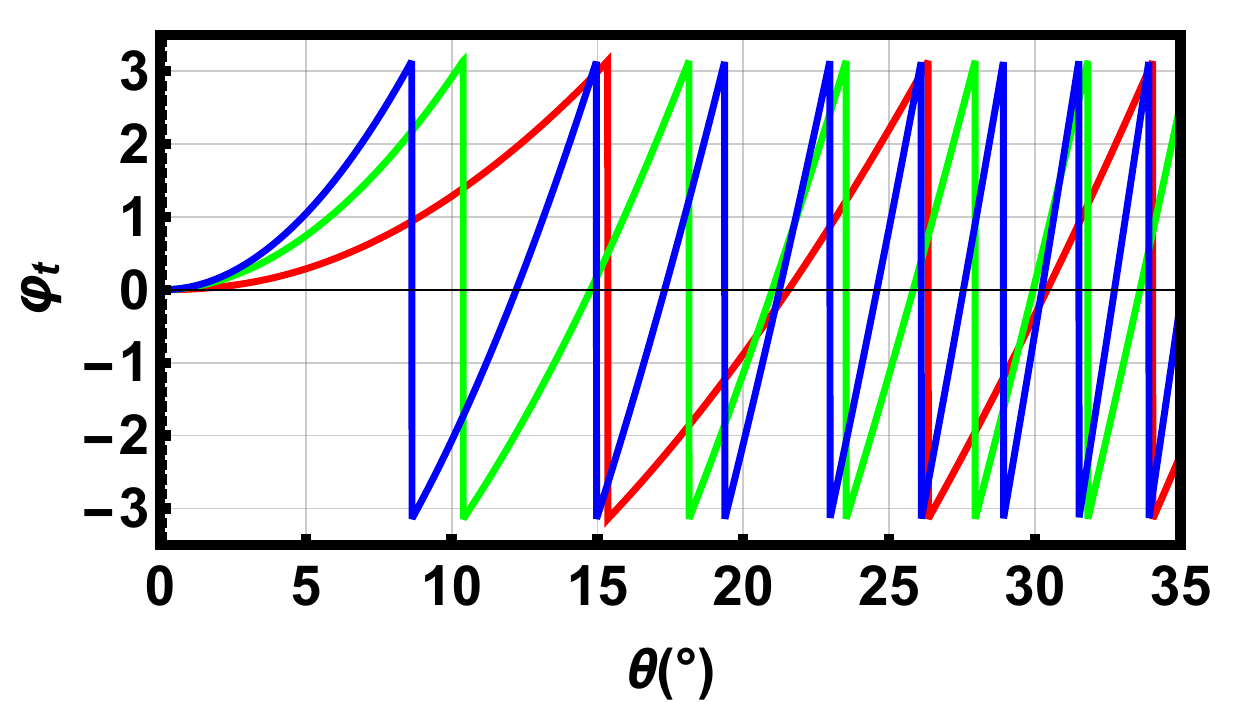}
        \label{FigPhiThetaE123piq13:B}
    }
    \subfloat[]{\centering
        \includegraphics[width=5.25cm]{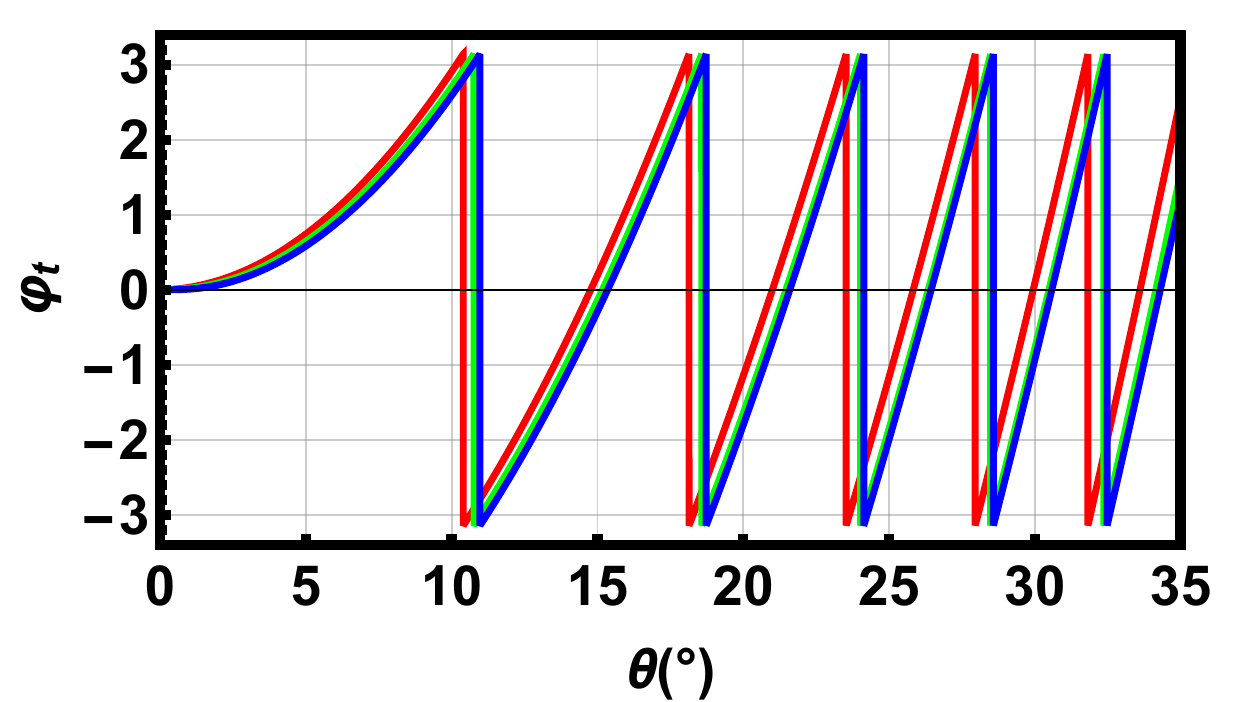}
        \label{FigPhiThetaE2piq13:C}
    }
     \caption{
     (color online) The phase shifts $\varphi_t$ versus the incident angle $\theta$ for the transmitted electron beam in graphene with spatially modulated potential of $n=30$ unit cells, for $\mathbb{V}=4\pi$. \protect\subref{FigPhiThetaE1pi:A}: $q_2=1/3$ for different values of incident energy near Dirac point $\varepsilon_D=\pi$ corresponding to red ($\varepsilon<\varepsilon_D$), green ($\varepsilon_D=\pi$) and blue ($\varepsilon>\varepsilon_D$) lines, respectively. \protect\subref{FigPhiThetaE123piq13:B}: $q_2=1/3$ for three values of incident energy $\varepsilon=\pi$ (red line), $2\pi$ (green line), $3\pi$ (blue line). 
     \protect\subref{FigPhiThetaE2piq13:C}: $\varepsilon=2\pi$ for three values of the distance $q_2=1/3$ (red line), $1/2$ (green line), $2/3$ (blue line). 
     }
     \label{PhaseShifts}
\end{figure}

Figure \ref{PhaseShifts} presents the phase shifts $\varphi_t$ versus the incident angle $\theta$ for the transmitted electron beam in graphene with spatially modulated potential of $n=30$ unit cells for  $\mathbb{V}=4\pi$ under some choices.  Figure \ref{PhaseShifts}\subref{FigPhiThetaE1pi:A} shows  $\varphi_t$ of the transmitted beam versus  $\theta$ with $q_2=1/3$ for different values of incident energy near the first Dirac point at finite energy $\varepsilon_D=\pi$ highlighted in  red ($\varepsilon<\varepsilon_D$), green ($\varepsilon=\varepsilon_D$) and blue ($\varepsilon>\varepsilon_D$) lines, respectively. We notice that, for $0\leq \theta\leq \theta_0$, $\varphi_t$ are positive and reach a positive peak for $\varepsilon\leq\varepsilon_D$, which is valid for $\varepsilon_D=m\pi$, $m$ is an integer as illustrated in Figure \ref{PhaseShifts}\subref{FigPhiThetaE123piq13:B}. For $\varepsilon>\varepsilon_D$ the phase shifts decrease from zero to negative regime and then reach the same value of positive peak. After $\theta_0$, they are oscillating periodically from positive to negative values with the same amplitudes and change the phase as long as  $\theta$ increases. In addition, the oscillation period decreases when $\theta$ increase. Figure \ref{PhaseShifts}\subref{FigPhiThetaE2piq13:C} shows  $\varphi_t$ of the transmitted beam versus  $\theta$ for different values of width $q_2$ where we observe that $\varphi_t$ vary smoothly with respect to  $q_2$.
\section*{Conclusion}\label{Conclusion}
To summarize, we have studied the  Goos-Hänchen shifts for Dirac fermions in graphene with spatially modulated potential. After setting the solutions of the energy spectrum, we have calculated the phase shifts of the transmitted and reflected beams via the transmission and reflection coefficients. These phases have been used to obtain the  Goos-Hänchen shifts in transmission and reflection as function of a set of physical
parameters characterizing our system such that the incident energy, potential height, incident angle and width of the central region of one cell.

Subsequently, we have presented different numerical results to emphasize the basic features of our system. Indeed, we have found that the  shifts near the Dirac points located at finite energy $\varepsilon_D=m\pi$, $m$ is an integer, depend strongly on the width $q_2$. Moreover, when $\varepsilon<\varepsilon_D$ the GH  shifts are always negative for any distance $q_2$, but it become positive when $\varepsilon>\varepsilon_D$. Furthermore, we have also found that at normal incidence the GH  shifts are null and as long as the incident angle increases the  the shifts increase. We have showed that, whatever $\varepsilon_D=m\pi$ ($m$ is an integer), for $0\leq \theta\leq \theta_0$, the phase shifts are positive and reach a positive peak for $\varepsilon\leq\varepsilon_D$. Contrariwise, for $\varepsilon>\varepsilon_D$ the phase shifts decrease from zero to negative regime and then reach the same value of positive peak. In case of  $\theta>\theta_0$, these phase shifts are oscillating periodically from positive to negative values with the same amplitudes and change the phase as long as incident angle $\theta$ increases.
We have noticed the GH  shifts  can be controlled by the incident energy, incident angle, the potential height as well as depend strongly (smoothly) on the width $q_2$ of central region of unit cell.
\section*{Acknowledgment}
The generous support provided by the Saudi Center for Theoretical Physics (SCTP) is highly appreciated by all authors.
\bibliographystyle{phjcp}
\bibliography{Bibliography_GH}
\end{document}